\documentclass[preprint,12pt,5p,times,twocolumn]{elsarticle}

\usepackage{epstopdf}
\usepackage{booktabs}
\usepackage{mathrsfs}
\usepackage{multirow}
\usepackage{amsmath,amsthm,amssymb,bm} % math
\usepackage{multirow}
\usepackage{hyperref}

\usepackage{dcolumn}% Align table columns on decimal point
\usepackage{longtable}
\usepackage[utf8]{inputenc}
\newcommand{\be}{\begin{eqnarray}}
\newcommand{\ee}{\end{eqnarray}}

\newcommand{\vslash}{{v\hspace{-5.4pt}/}}

\usepackage[normalem]{ulem}  % \sout{old text} for strikeout
\usepackage[dvips]{color} % For blue in-text comments and additions
\renewcommand\sout{\bgroup \color{red} \ULdepth=-.5ex \ULset}

\biboptions{sort&compress}

\begin{document}

\begin{frontmatter}



\title{
Hidden-charm pentaquarks as a meson-baryon molecule 
with coupled channels for $\bar{D}^{(\ast)}\Lambda_{\rm c}$ and
$\bar{D}^{(\ast)}\Sigma^{(\ast)}_{\rm c}$
}
%\tnotetext[a]{Report No.: KEK-TH-1676, J-PARC-TH-29}


\author[a]{Yasuhiro Yamaguchi\corref{cor1}}
\ead{yasuhiro.yamaguchi@ge.infn.it}
\cortext[cor1]{Corresponding author.}
\author[a]{Elena Santopinto}
\ead{elena.santopinto@ge.infn.it}

\address[a]{Istituto Nazionale di Fisica Nucleare (INFN), Sezione di Genova, via Dodecaneso 33, 16146 Genova, Italy}

\begin{abstract}

 The recent observation of two hidden-charm pentaquark states by LHCb
 collaborations prompted
 us to
 investigate the exotic states close to the 
 $\bar{D}\Lambda_{\rm c}$,
 $\bar{D}^{\ast}\Lambda_{\rm c}$,
 $\bar{D}\Sigma_{\rm c}$,
 $\bar{D}\Sigma^{\ast}_{\rm c}$,
 $\bar{D}^{\ast}\Sigma_{\rm c}$ and
 $\bar{D}^{\ast}\Sigma^{\ast}_{\rm c}$
 thresholds.
 %$\bar{D}^{(\ast)}\Sigma^{(\ast)}_{\rm c}$ thresholds.
 We therefore studied the hadronic molecules that form the
 coupled-channel
 system of
 $\bar{D}^{(\ast)}\Lambda_{\rm c}$ and
 $\bar{D}^{(\ast)}\Sigma^{(\ast)}_{\rm c}$.
 As the heavy quark spin symmetry 
 manifests the mass degenerations of $\bar{D}$ and $\bar{D}^\ast$ mesons,
 and of $\Sigma_{\rm c}$ and $\Sigma^\ast_{\rm c}$ baryons, 
 the coupled channels of
 $\bar{D}^{(\ast)}\Sigma^{(\ast)}_{\rm c}$ 
 are important in these
 molecules.
 In addition, we consider the coupling to the $\bar{D}^{(\ast)}\Lambda_{\rm c}$
 channel whose thresholds are near 
 the $\bar{D}^{(\ast)}\Sigma^{(\ast)}_{\rm c}$ thresholds,
 and the coupling to the state with nonzero orbital angular momentum
 mixed by the tensor force.
 This full coupled channel analysis of 
 $\bar{D}^{(\ast)}\Lambda_{\rm c}-\bar{D}^{(\ast)}\Sigma^{(\ast)}_{\rm
 c}$ with larger orbital angular momentum 
 has never been performed before.
 By solving the coupled-channel Schr\"odinger equations with
 the one meson exchange potentials that respected to the heavy
 quark spin and chiral symmetries,
 we studied
 the hidden-charm hadronic
 molecules with $I(J^P)=1/2(3/2^\pm)$ and $1/2(5/2^\pm)$.
 We conclude that the $J^P$ assignment of the observed pentaquarks is
 $3/2^+$ for $P^+_{\rm c}(4380)$ and $5/2^-$ for $P^+_{\rm c}(4450)$,
 which is agreement with the results of the LHCb analysis.
 In addition, we give predictions for other $J^P=3/2^\pm$ states
 at 4136.0, 4307.9 and 4348.7 MeV in $J^P=3/2^-$, and 4206.7 MeV in $J^P=3/2^+$, which
 can be further investigated by means of experiment.
\end{abstract}


\end{frontmatter}


%\section{}
%\label{}

%======================================================
\section{Introduction}
%======================================================
In 2015, LHCb collaborations reported the two hidden-charm pentaquarks
$P^+_{\rm c}(4380)$ and $P^+_{\rm c}(4450)$
in $\Lambda^0_{\rm b}\rightarrow J/\psi K^- p$ decay~\cite{Aaij:2015tga,Aaij:2016phn,Aaij:2016ymb}.
The reported masses and widths are 
$(M,\Gamma)=(4380\pm 8\pm 29, 205\pm 18\pm 86)$ MeV and 
$(4449.8\pm 1.7\pm 2.5, 39\pm 5\pm 19)$ MeV, respectively,
which are close to $\bar{D}\Sigma^\ast_{c}$ and
$\bar{D}^\ast\Sigma_{c}$ thresholds.
Their significances are 9 and 12 standard deviations,
respectively.
The total angular momentum is $3/2$ for one state and $5/2$ for the
other.
These states have opposite parity.
The minimal quark content of the pentaquarks is $c\bar{c}uud$ because
the states decay into $J/\psi p$.
In the literature there have been lively discussion about the
structure of the hidden-charm pentaquarks.
The compact pentaquark states
have been discussed by the (di)quark
model~\cite{Wang:2011rga,Yuan:2012wz,Maiani:2015vwa,Li:2015gta,Yang:2015bmv}
and G\"ursey-Radicati inspired formula~\cite{Santopinto:2016pkp}.
The hadronic molecules have been studied by 
the meson-baryon coupled-channel 
approach~\cite{Hofmann:2005sw,PhysRevLett.105.232001,PhysRevC.84.015202,
Garcia-Recio:2013gaa,Xiao:2013yca,Karliner:2015ina,Uchino:2015uha,Chen:2016heh,Shimizu:2016rrd}
and 
the QCD sum rules~\cite{Chen:2015moa,Wang:2015epa}.
On the other hand, the threshold enhancement by the anomalous triangle
singularity is discussed in Refs.~\cite{Guo:2015umn,Liu:2015fea,Guo:2016bkl}.

Near the thresholds, resonances are expected to have the
exotic structure, like the hadronic molecule.
In the strangeness sector, 
$\Lambda(1405)$ is considered to be generated
by the $\bar{K}N$ and $\pi\Sigma$~\cite{Oset:1997it,Krippa:1998us,Hyodo:2011ur}.
In the heavy quark sectors,
$X(3872)$ has the dominant component of the $D\bar{D}^\ast$
molecules~\cite{PhysRevLett.91.262001,Swanson:2006st,Brambilla:2010cs,Gamermann:2007fi,Ferretti:2013faa}.
The charged quarkonium states
$Z_c(3900)$~\cite{Ablikim:2013mio} and 
$Z^{(\prime)}_b$~\cite{PhysRevLett.108.122001}
are considered to be
$D\bar{D}^\ast$~\cite{Wang:2013cya} and $B^{(\ast)}\bar{B}^\ast$~\cite{Bondar:2011ev,Ohkoda:2011vj,Hosaka:2016pey},
respectively.
The observed pentaquarks are found just below the $\bar{D}\Sigma^\ast_{\rm c}$
and $\bar{D}^\ast\Sigma_{\rm c}$ thresholds.
Therefore the $\bar{D}\Sigma^\ast_{\rm c}$ and $\bar{D}^\ast\Sigma_{\rm
c}$ molecular components are expected to be dominant.

Since the hadronic molecules are dynamically generated by the
hadron-hadron interaction,
the properties of the interaction are important in producing those
structures.
In the literature,
the SU(4) flavor symmetric interaction has been applied to the charm sector.
This is an extension of the interaction based on the SU(3) flavor symmetry
applied to the strangeness sector.
In the hidden-charm pentaquarks, the interactions based on the SU(4)
flavor symmetry have been
used~\cite{Hofmann:2005sw,PhysRevLett.105.232001,PhysRevC.84.015202}.
However, the SU(4) symmetry is expected to be broken because the mass of
the charm quark is much larger than those of light quarks.

In the heavy flavor sector,
new symmetry of heavy quarks emerges which is called heavy quark
symmetry~\cite{Isgur:1989vq,Isgur:1989ed,PhysRevLett.66.1130,ManoharWise200707}.
This results from
the suppression of the spin-dependent interaction
among heavy quarks.
It manifests the mass degeneracy of the
states with different total spin, e.g. degeneracies of
$D$ and $D^\ast$ mesons ($\Delta m_{DD^\ast}\sim 140$ MeV) and
$\Sigma_c$ and $\Sigma^\ast_c$ baryons ($\Delta m_{\Sigma_{\rm c}\Sigma_{\rm c}^\ast}\sim 65$ MeV).
Therefore, hadronic states should be a coupled-channel system.
In that case, 
thresholds of $\bar{D}^{(\ast)}\Sigma^{(\ast)}_{\rm c}$
($\bar{D}^{(\ast)}=\bar{D}$ or $\bar{D}^{\ast}$, and $\Sigma^{(\ast)}_{\rm
c}=\Sigma_{\rm c}$ or $\Sigma_{\rm c}^{\ast}$)
are close to the states we are going to study
(see also Table~\ref{table:coupledchannel12}).

Moreover, we cannot ignore the 
$\bar{D}^{(\ast)}\Lambda_{\rm c}$
channel.
In the strangeness sector, the $\Lambda-\Sigma$
mixing is important in the hyperon-nucleon
interaction~\cite{PhysRevC.40.R7}.
In the early
works~\cite{PhysRevLett.105.232001,PhysRevC.84.015202,Chen:2016heh,Shimizu:2016rrd}, 
however,
the coupling to $\bar{D}^{(\ast)}\Lambda_{\rm c}$
is not considered in the hidden-charm pentaquarks.
However, 
the $\bar{D}^{\ast}\Lambda_{\rm c}$ threshold is
25 MeV below the $\bar{D}\Sigma_{\rm c}$ threshold.
Therefore, the $\bar{D}^{\ast}\Lambda_{\rm c}$ channel is not
irrelevant in the hidden-charm meson-baryon molecules.

The approximate mass degeneracy of heavy hadrons
changes the aspect of interactions in the heavy quark sector.
Indeed, the $\bar{D}N-\bar{D}^\ast N$
mixing enhances the effect of the one pion exchange potential (OPEP)
in the $\bar{D}$ meson-nucleon ($\bar{D}N$) system,
while the $KN-K^\ast N$ mixing is suppressed due to the large mass
difference between $K$ and $K^\ast$ mesons ($\Delta m_{KK^\ast}\sim 400$
MeV) in the strangeness sector.
In nuclear physics, the OPEP is the basic ingredient of the nuclear force
that binds the atomic nuclei.
Specifically, the tensor force mixing $S$-wave and $D$-wave components
yields the strong attraction.
This mechanism has been suggested to have an important role in the 
$\bar{D}^{(\ast)} N$ system 
in
Refs.~\cite{Yasui:2009bz,PhysRevD.84.014032,PhysRevD.85.054003,Yamaguchi:2013ty,PhysRevD.72.074010,Gamermann:2010zz,Liang:2014kra,Hosaka:2016ypm}.
The coupled-channel analysis with the mixing of $S$-wave and $D$-wave
was not performed in
Refs.~\cite{PhysRevLett.105.232001,PhysRevC.84.015202}.
However,
this mixing is helpful to produce the attraction in the hidden-charm
molecules.

On the basis of the above discussions,
we consider the coupled-channel systems 
of $\bar{D}^{(\ast)}\Lambda_c-\bar{D}^{(\ast)}\Sigma^{(\ast)}_c$
including states with larger orbital angular momentum,
 namely $D$-wave and $G$-wave for the negative parity state and 
 $F$-wave and $H$-wave for the positive parity state, as summarized in
Table~\ref{table:coupledchannel12}.
This full-channel coupling
has never been considered so far. 
The interaction is obtained by the one meson exchange potential
that respects the heavy quark spin symmetry.
The bound and resonant states in $I(J^P)=1/2(3/2^\pm)$ and
$1/2(5/2^\pm)$ are studied by solving the coupled-channel
Schr\"odinger equations. 
In this study, the $J/\psi N$ channel is not considered 
because the coupling to the $J/\psi N$ channel with the charmed meson
exchange would be suppressed and the molecular state is dominated by the 
$\bar{D}^{(\ast)}\Lambda_c$ and $\bar{D}^{(\ast)}\Sigma^{(\ast)}_c$ channels.

This paper is organized as follows.
The meson exchange potentials between the charmed meson and
baryon are shown in Sec.~\ref{Sec:Interactions}.
The numerical results are summarized in
Sec.~\ref{Sec:result}.
Sec.~\ref{Sec:Summary} 
summarizes the work as a whole.

%========================================%
\section{Interactions}
\label{Sec:Interactions}
%========================================%

%========================================%
% Table: Channels
%========================================%
\begin{table*}[t] %[htbp]
 \caption{Various channels of the $\bar{D}^{(\ast)}\Lambda_{\rm c}$ and
 $\bar{D}^{(\ast)}\Sigma^{(\ast)}_{\rm c}$ states for
 the given total angular momentum and parity $J^P$ 
 and the corresponding thresholds.
 In each channel, the total spin $S$ and orbital angular momentum $L$ is
 represented as $^{2S+1}L$.
 The thresholds as a sum of the mass of the meson and baryon in the last row
 are given in the unit of MeV.
 }
 \label{table:coupledchannel12}
 \begin{center}
  \begin{tabular}{c|l}
   \hline\hline
   $J^P$&Channels \\
   \hline
   $3/2^-$&$\bar{D}\Lambda_{\rm c}(^2D),\,\,
       \bar{D}^\ast\Lambda_{\rm c}(^4S,^2D,^4D),\,\,
       \bar{D}\Sigma_{\rm c}(^2D),\,\,
       \bar{D}\Sigma^\ast_{\rm c}(^4S,^4D),\,\,
       \bar{D}^\ast\Sigma_{\rm c}(^4S,^2D,^4D),\,\,
       \bar{D}^\ast\Sigma^\ast_{\rm c}(^4S,^2D,^4D,^6D,^6G)$ \\
   $3/2^+$&$\bar{D}\Lambda_{\rm c}(^2P),\,\,
       \bar{D}^\ast\Lambda_{\rm c}(^2P,^4P,^4F),\,\,
       \bar{D}\Sigma_{\rm c}(^2P),\,\,
       \bar{D}\Sigma^\ast_{\rm c}(^4P,^4F),\,\,
       \bar{D}^\ast\Sigma_{\rm c}(^2P,^4P,^4F),\,\,
       \bar{D}^\ast\Sigma^\ast_{\rm c}(^2P,^4P,^6P,^4F,^6F)$ \\
   $5/2^-$&$\bar{D}\Lambda_{\rm c}(^2D),\,\,
       \bar{D}^\ast\Lambda_{\rm c}(^2D,^4D,^4G),\,\,
       \bar{D}\Sigma_{\rm c}(^2D),\,\,
       \bar{D}\Sigma^\ast_{\rm c}(^4D,^4G),\,\,
       \bar{D}^\ast\Sigma_{\rm c}(^2D,^4D,^4G),\,\,
       \bar{D}^\ast\Sigma^\ast_{\rm c}(^6S,^2D,^4D,^6D,^4G,^6G)$ \\
   $5/2^+$&$\bar{D}\Lambda_{\rm c}(^2F),\,\,
       \bar{D}^\ast\Lambda_{\rm c}(^4P,^2F,^4F),\,\,
       \bar{D}\Sigma_{\rm c}(^2F),\,\,
       \bar{D}\Sigma^\ast_{\rm c}(^4P,^4F),\,\,
       \bar{D}^\ast\Sigma_{\rm c}(^4P,^2F,^4F),\,\,
       \bar{D}^\ast\Sigma^\ast_{\rm c}(^4P,^6P,^2F,^4F,^6F,^6H)$ \\
   \hline\hline
   &Thresholds (MeV) \\ \hline
   &$\bar{D}\Lambda_{\rm c}(4153.5),\,\,
       \bar{D}^\ast\Lambda_{\rm c}(4295.5),\,\,
       \bar{D}\Sigma_{\rm c}(4320.5),\,\,
       \bar{D}\Sigma^\ast_{\rm c}(4385.1),\,\,
       \bar{D}^\ast\Sigma_{\rm c}(4462.5),\,\,
       \bar{D}^\ast\Sigma^\ast_{\rm c}(4527.1)$
   \\
   \hline\hline
  \end{tabular}
 \end{center}
\end{table*}

The Lagrangians satisfying the heavy quark and chiral symmetries are
employed.
The Lagrangians for a heavy meson and a light meson are given~\cite{PhysRevD.45.R2188,Falk:1992cx,PhysRevD.46.1148,Casalbuoni:1996pg,ManoharWise200707} as
\begin{align}
 {\cal L}_{\pi HH}=&g_{\pi}{\rm Tr}\left[
 H_b\gamma_\mu\gamma_5A^\mu_{ba}\bar{H}_a\right] \, ,\\
 {\cal L}_{v HH}=&-i\beta{\rm Tr}\left[
 H_bv^\mu(\rho_\mu)_{ba}\bar{H}_a \right] \notag \\
 &+i\lambda
 {\rm Tr}\left[H_b\sigma^{\mu\nu}F_{\mu\nu}(\rho)_{ba}\bar{H}_a\right] 
 \,,\\
 {\cal L}_{\sigma HH}=&g_s{\rm Tr}\left[H_a\sigma\bar{H}_a\right] \, ,
\end{align}
where the subscripts $\pi$, $v$ and $\sigma$ are for the pion, vector
meson ($\rho$ and $\omega$) and sigma meson, respectively.
$v^\mu$ is a four-velocity of a heavy quark.
The heavy meson field constructed by the pseudoscalar meson $P$ and vector
meson $P^\ast$ are represented~\cite{PhysRevD.45.R2188,Falk:1992cx,PhysRevD.46.1148,Casalbuoni:1996pg,ManoharWise200707} by 
\begin{align}
 H_a&=\frac{1+\vslash}{2}\left[P^\ast_{a\mu}\gamma^\mu-P_a\gamma_5
 \right] \, ,\\
 \bar{H}_a&=\gamma_0H^\dagger_a\gamma_0\, ,
\end{align}
where the subscripts $a,b$ are for the light flavor.
The axial vector current $A_\mu$ is given by
\begin{align}
 A_\mu&=\frac{i}{2}\left(\xi^\dagger\partial_\mu\xi-\xi\partial_\mu\xi^\dagger\right),
\end{align}
where $\xi=\exp(i\hat{\pi}/2f_\pi)$ with the pion decay constant
$f_\pi=92.3$ MeV.
The pseudoscalar and vector meson fields are given by
\begin{align}
 \hat{\pi}&=\sqrt{2}\left(
 \begin{array}{ccc}
  \frac{\pi^0}{\sqrt{2}}+\frac{\eta}{\sqrt{6}}&\pi^+&K^+\\
  \pi^-&-\frac{\pi^0}{\sqrt{2}}+\frac{\eta}{\sqrt{6}}&K^0\\
  K^-&\bar{K}^0&-\frac{2}{\sqrt{6}}\eta\\
 \end{array}
 \right) \, ,\\
 \rho_\mu&=i\frac{g_V}{2}\hat{\rho}_\mu \, ,\\
  \hat{\rho}_\mu&=\sqrt{2}\left(
 \begin{array}{ccc}
  \frac{\rho^0}{\sqrt{2}}+\frac{\omega}{\sqrt{2}}&\rho^+&K^{\ast\,+}\\
  \rho^-&-\frac{\rho^0}{\sqrt{2}}+\frac{\omega}{\sqrt{2}}&K^{\ast\,0}\\
  K^{\ast\,-}&\bar{K}^{\ast\,0}&\phi\\
 \end{array}
 \right)_\mu \, , \\
 F_{\mu\nu}(\rho)&=\partial_\mu\rho_\nu-\partial_\nu\rho_\mu \, .
\end{align}
The gauge coupling constant $g_V$ is obtained as
$g_V=m_\rho/\sqrt{2}f_\pi$~\cite{Bando:1987br}.

The $\pi PP^\ast$ coupling constant is determined by the strong decay of
$D^\ast \rightarrow D\pi$~\cite{Agashe:2014kda}.
The coupling constants $\beta$ and $\lambda$ are fixed by the vector
meson decays~\cite{PhysRevD.68.114001}.
The coupling constant for the sigma meson is given by $g_s=-g^\prime_\pi/2\sqrt{6}$ with the $0^+\rightarrow 0^-\pi$ coupling
constant $g^\prime_\pi=3.73$~\cite{PhysRevD.68.054024}.
These coupling constants are summarized in Table~\ref{table:couplings}.

%========================================%
% Table: Coupling constants
%========================================%
\begin{table*}[t]
 \caption{Masses of the exchanged mesons and coupling constants of the interaction Lagrangians for the
 heavy mesons and heavy baryons~\cite{PhysRevD.84.014032,PhysRevD.68.114001,PhysRevD.68.054024,Liu:2011xc}.}
 \label{table:couplings}
 \begin{center}
  \begin{tabular}{cccc}
   \hline\hline
   &$m_\alpha$ [MeV] &Meson &Baryon \\ \hline
   $\pi$&137.27 &$g_\pi=0.59$ &$g_1=(3/\sqrt{8})g_4=1.0$ \\ 
   $\rho$&769.9 &$\beta=0.9$
	   %(\beta,\lambda)=(0.9,0.59\, \text{[GeV$^{-1}$]})$ 
	   &$\beta_S=-2\beta_B=12.0/g_V$
	       % $(\beta_S,\lambda_S)=(12.0/g_V,19.2/g_V\,
	       % \text{[GeV$^{-1}$]})$ 
	       \\
   &&$\lambda=0.59\, \text{[GeV$^{-1}$]}$& 
       $\lambda_S=-2\sqrt{2}\lambda_I=19.2/g_V\,\text{[GeV$^{-1}$]}$\\
   $\omega$&781.94 &$\beta=0.9$
	   %(\beta,\lambda)=(0.9,0.59\, \text{[GeV$^{-1}$]})$ 
	   &$\beta_S=-2\beta_B=12.0/g_V$
	       % $(\beta_S,\lambda_S)=(12.0/g_V,19.2/g_V\,
	       % \text{[GeV$^{-1}$]})$ 
	       \\
	   %$(\beta,\lambda)=(0.9,0.59\, \text{[GeV$^{-1}$]})$ 
%	   & $(\beta_S,\lambda_S)=(12.0/g_V,19.2/g_V\,
       %	   \text{[GeV$^{-1}$]})$\\
   &&$\lambda=0.59\, \text{[GeV$^{-1}$]}$& 
       $\lambda_S=-2\sqrt{2}\lambda_I=19.2/g_V\,\text{[GeV$^{-1}$]}$\\
   $\sigma$&550.0  &$g_s=-0.76$ &$\ell_S=-2\ell_B=7.30$ \\
   \hline\hline
  \end{tabular}
 \end{center}
\end{table*}

The Lagrangians for a heavy baryon and a light meson~\cite{PhysRevD.46.1148,Liu:2011xc} are given by
\begin{align}
 {\cal L}_{\pi BB}=&
  \frac{3}{2}g_1 iv_\kappa\epsilon^{\mu\nu\lambda\kappa}{\rm tr}
 \left[\bar{S}_\mu A_\nu S_\lambda \right]
 +g_4{\rm tr}\left[\bar{S}^\mu A_\mu B_{\bar{3}}\right] + {\rm H.c.} \, , \\
 {\cal L}_{vBB}=&-i\beta_B{\rm tr}\left[
 \bar{B}_{\bar{3}}v^\mu\rho_\mu B_{\bar{3}}
 \right]
 -i\beta_S{\rm tr}\left[
 \bar{S}_\mu v^\alpha\rho_\alpha S^\mu
 \right]
  \notag \\
 &+\lambda_S{\rm tr}\left[
 \bar{S}_\mu F^{\mu\nu} S_\nu
 \right]
 +i\lambda_I\epsilon^{\mu\nu\lambda\kappa}v_\mu{\rm tr}\left[
 \bar{S}_\nu F_{\lambda\kappa}B_{\bar{3}}
 \right] + {\rm H.c.} \, , \\
 {\cal L}_{\sigma BB}=&\ell_{B}
 {\rm tr}\left[\bar{B}_{\bar{3}}\sigma B_{\bar{3}}\right]
 +\ell_{S}
 {\rm tr}\left[\bar{S}_{\mu}\sigma S^{\mu}\right] \, .
\end{align}
The superfield $S_\mu$ for $\Sigma^{(\ast)}_Q$ is represented by
\begin{align}
 S_\mu=&B^\ast_{6\mu}+\frac{\delta}{\sqrt{3}}\left(
 \gamma_\mu+v_\mu\right)\gamma_5B_6 \, , \\
 \bar{S}_\mu=&\gamma_0S^\dagger_\mu\gamma_0 \, .
\end{align}
The phase factor is chosen by $\delta=-1$ as discussed in
Ref~\cite{Liu:2011xc}.
Heavy baryon fields are expressed by the $3\times 3$ matrix
form~\cite{PhysRevD.46.1148,Liu:2011xc};
\begin{align}
 B_6&=\left(
 \begin{array}{ccc}
  \Sigma^{+1}_Q&\frac{1}{\sqrt{2}}\Sigma^{0}_Q
   &\frac{1}{\sqrt{2}}\Xi^{\prime +1/2}_Q \\
  \frac{1}{\sqrt{2}}\Sigma^{0}_Q&\Sigma^{-1}_Q &\frac{1}{\sqrt{2}}\Xi^{\prime -1/2}_Q \\
  \frac{1}{\sqrt{2}}\Xi^{\prime +1/2}_Q& \frac{1}{\sqrt{2}}\Xi^{\prime
   -1/2}_Q & \Omega_Q \\
 \end{array}
 \right)\,  , \\
 B_{\bar{3}}&=\left(
 \begin{array}{ccc}
  0&\Lambda_Q &\Xi^{+1/2}_Q \\
  \Lambda_Q& 0&\Xi^{-1/2}_Q \\
  -\Xi^{+1/2}_Q& -\Xi^{\prime -1/2}_Q& 0\\
 \end{array}
 \right) \, .
\end{align}
The matrix for $B^\ast_6$ is similar to $B_6$.
The field of the $B^\ast_6$ baryon is given by the Rarita-Schwinger
field~\cite{Rarita:1941mf,Liu:2011xc}.
We use the coupling constants given by the quark model estimation discussed
in Ref.~\cite{Liu:2011xc}.

From the effective Lagrangians introduced above,
we obtain the meson exchange potentials as
\begin{align}
 V^{ij}_{\pi}(r)&= G^{ij}_\pi\left[\vec{\cal O}^i_1\cdot\vec{\cal O}^j_2C(r;m_\pi)
 +S_{{\cal O}^i_1{\cal O}^j_2}(\hat{r})T(r;m_\pi)\right] \, , 
 \label{eq:pionpote} \\
 V^{ij}_{v}(r)&= G^{ij}_{v}C(r;m_v) \notag\\
 &+F^{ij}_v\left[
 -2\vec{\cal O}^i_1\cdot\vec{\cal O}^j_2C(r;m_v)
 +S_{{\cal O}^i_1{\cal O}^j_2}(\hat{r})T(r;m_v)
 \right] \, , 
 \label{eq:vectorpote} \\
 V^{ij}_{\sigma}(r)&= G^{ij}_{\sigma}C(r;m_\sigma) \, .
 \label{eq:sigmapote}
\end{align}
In this study, we suppress the potential term, which is proportional to 
the inverse of the heavy baryon mass.
$i$ and $j$ stand for the indices of the channels.
$G^{ij}_\alpha$ ($\alpha=\pi,\rho,\omega,\sigma$) 
is the constant of the ($i,j$) component 
given by the coupling constants of the Lagrangians.
${\cal O}^i_1$ and ${\cal O}^j_2$ are the (transition) spin operator 
of the heavy meson and heavy baryon vertices, respectively~\cite{Yasui:2009bz,PhysRevD.84.014032,PhysRevD.85.054003,Liu:2011xc}.
$S_{{\cal O}^i_1{\cal O}^j_2}(\hat{r})$ is the tensor operator $S_{{\cal
O}^i_1{\cal O}^j_2}(\hat{r})=3\vec{\cal O}^i_1\cdot\hat{r}\,
\vec{\cal O}^j_2\cdot\hat{r}-\vec{\cal O}^i_1\cdot\vec{\cal
O}^j_2$.
The potential for the isovector mesons, $\pi$ and $\rho$,
is multiplied by the isospin factor, $\sqrt{6}$ for
$\bar{D}^{(\ast)}\Lambda_{\rm c}-\bar{D}^{(\ast)}\Sigma^{(\ast)}_{\rm c}$
and $-2$ for $\bar{D}^{(\ast)}\Sigma^{(\ast)}_{\rm
c}-\bar{D}^{(\ast)}\Sigma^{(\ast)}_{\rm c}$ 
with $I=1/2$.
The functions $C(r;m_\alpha)$ and $T(r;m_\alpha)$ are given by
\begin{align}
 &C(r;m_\alpha)=\int\frac{d^3q}{(2\pi)^3}
 \frac{m^2_\alpha}{\vec{q}\,^2+m^2_\alpha}
 e^{i\vec{q}\cdot\vec{r}}F_\alpha (\Lambda,\vec{q})
 \, ,\\
 &S_{{\cal O}^i_1{\cal O}^j_2}(\hat{r})T(r;m_\alpha) \notag\\
 &=\int\frac{d^3q}{(2\pi)^3}
 \frac{-\vec{q}\,^2}{\vec{q}\,^2+m^2_\alpha}
 S_{{\cal O}^i_1{\cal O}^j_2}(\hat{q})e^{i\vec{q}\cdot\vec{r}}F_\alpha(\Lambda,\vec{q})\, .
\end{align}
We introduce the standard dipole-type form factor $F(\Lambda,\vec{q})$ for spatially extended
hadrons~\cite{Yasui:2009bz,PhysRevD.84.014032,PhysRevD.85.054003,Liu:2011xc,Shimizu:2016rrd}
\begin{align}
 F_\alpha(\Lambda,\vec{q}\,)=&
 \frac{\Lambda^2_P-m^2_\alpha}{\Lambda^2_P+\vec{q}\,^2}
 \frac{\Lambda^2_B-m^2_\alpha}{\Lambda^2_B+\vec{q}\,^2} \, ,
\end{align}
with the cutoff parameters $\Lambda_P$ and $\Lambda_B$ for the heavy
meson and the heavy baryon, respectively.
In this study, we employ the common cutoff parameter
$\Lambda=\Lambda_P=\Lambda_B$ for simplicity,
as discussed in Refs.~\cite{Liu:2011xc,Shimizu:2016rrd}.
In this study, 
only the cutoff $\Lambda$ is a free parameter.
We determine $\Lambda$ in order to reproduce the mass spectra of 
the observed pentaquarks.

%========================================%
\section{Numerical results}
\label{Sec:result}
%========================================%

\begin{figure*}[t]
 \begin{center}
  \begin{tabular}{cc}
   (i) $I(J^P)=1/2(3/2^-)$& (ii) $I(J^P)=1/2(3/2^+)$ \\
   \includegraphics[width=0.45\linewidth,clip]{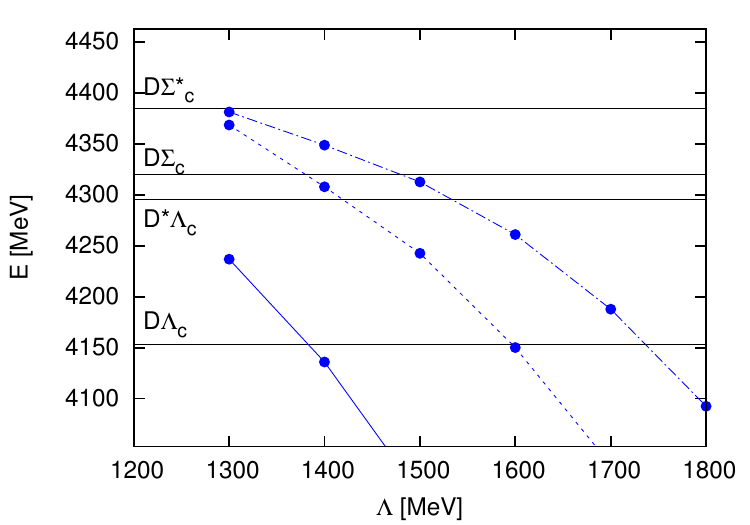}&
   \includegraphics[width=0.45\linewidth,clip]{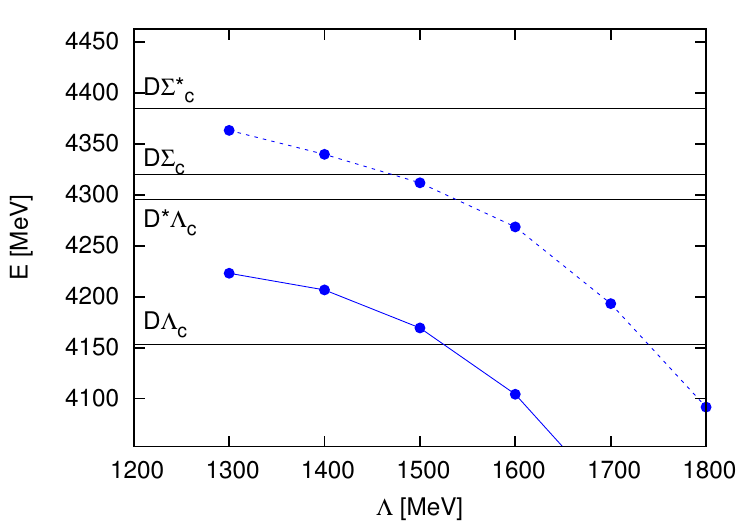} \\[2mm]
   (iii) $I(J^P)=1/2(5/2^-)$& (iv) $I(J^P)=1/2(5/2^+)$ \\
   \includegraphics[width=0.45\linewidth,clip]{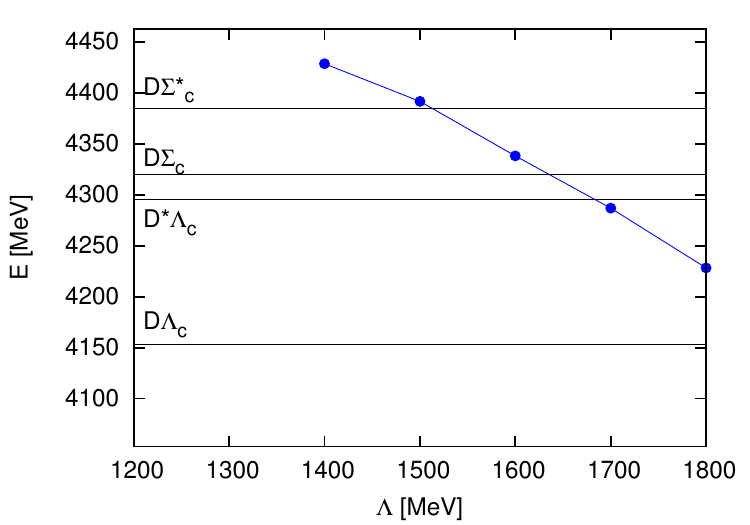}&
   \includegraphics[width=0.45\linewidth,clip]{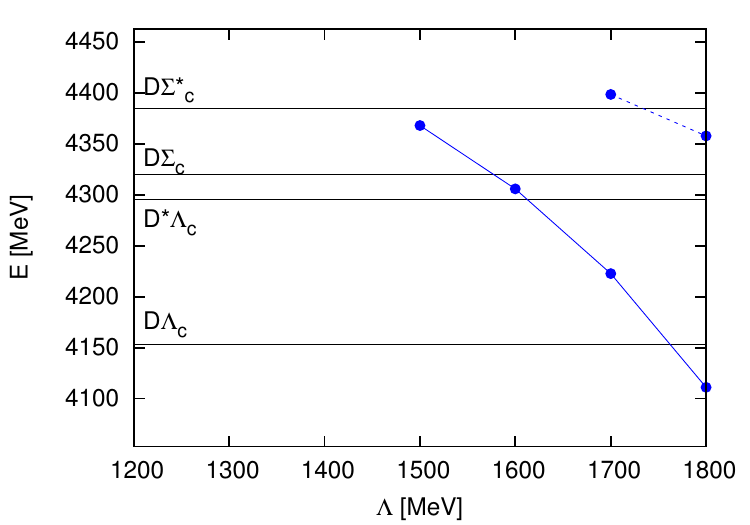} \\
  \end{tabular}
  \caption{Cutoff $\Lambda$ dependence of the obtained energies $E$ of
  in $J^P=3/2^\pm$ and $5/2^\pm$ and isospin $I=1/2$.
  The explicit values are shown in Table~\ref{table:resultsPB}.
  The black horizontal lines show the meson-baryon thresholds.}
  \label{fig:energyspectrum}
 \end{center}
\end{figure*}

\begin{table*}[t]
 \begin{center}
  \caption{Obtained energies 
  in $J^P=3/2^\pm$ and $5/2^\pm$
  with
  the various cutoffs $\Lambda$.
  The real energy gives the binding energy when 
  the value of $\bar{D}\Lambda_{\rm c}$ threshold ($= 4153.5$ MeV) is 
  subtracted.
  The complex energy is given by $E=E_{\rm re}-i\Gamma/2$ with
  the resonance energy $E_{\rm re}$ and the decay width $\Gamma$.
  Note that the decay to $J/\psi p$ is not considered in this study.
  }
  \label{table:resultsPB}
  \begin{tabular}{c|cccccc}
   \hline\hline
   $\Lambda$ [MeV]&1300&1400&1500&1600&1700&1800\\ \hline
   $J^P=3/2^-$&$4236.9-i0.8$&$4136.0$&$4006.3$&$3848.2$&$3660.0$&$3438.26$\\
              &$4381.3-i11.4$&$4307.9-i18.8$&$4242.6-i1.4$&$4150.1$&$4035.2$&$3897.3$\\
              &$4368.5-i64.9$&$4348.7-i21.1$&$4312.7-i16.0$&$4261.0-i7.0$&$4187.7-i0.9$&$4092.5$\\
   \hline   
   $J^P=3/2^+$ &$4223.0-i97.9$&$4206.7-i41.2$&$4169.3-i5.3$&$4104.2$&$3996.7$&$3855.8$\\
               &$4363.3-i57.0$&$4339.7-i26.8$&$4311.8-i6.6$&$4268.5-i1.3$&$4193.2-i0.1$&$4091.6$\\
   \hline
   $J^P=5/2^-$   &---&$4428.6-i89.1$&$4391.7-i88.8$&$4338.2-i56.2$&$4286.8-i27.3$&$4228.3-i7.4$\\
   \hline
   $J^P=5/2^+$ &---&---&$4368.0-i9.2$&$4305.8-i1.9$&$4222.7-i1.4$&$4111.1$\\
               &---&---&---&---&$4398.5-i15.0$&$4357.8-i8.2$\\
   \hline\hline
  \end{tabular}
 \end{center}
\end{table*}

The total Hamiltonian is given by the sum of the kinetic term and the
meson exchange potential between the heavy meson and the heavy baryon
for the coupled-channels in
Eqs.~\eqref{eq:pionpote}-\eqref{eq:sigmapote}.
The interaction is the heavy quark spin symmetric.
However,
the breaking effect of the symmetry is given by the
mass splittings of $\bar{D}$ and $\bar{D}^\ast$, and 
$\Sigma_{\rm c}$ and $\Sigma^\ast_{\rm c}$
in this calculation.
By diagonalizing the Hamiltonian, we obtain the energy of the bound and
resonant states.
The wave function is expressed by the Gaussian expansion
method~\cite{Hiyama:2003cu}.
In order to obtain a complex energy of a resonance,
the complex scaling method is used in this study~\cite{Aguilar1971_269,
Blaslev1971_22,Simon1972_27,Prog.Theor.Phys11620061Aoyama}.

We study the molecular states of 
$\bar{D}^{(\ast)}\Lambda^{(\ast)}_{\rm
c}-\bar{D}^{(\ast)}\Sigma^{(\ast)}$ with $J^P=3/2^\pm,5/2^\pm$ and
isospin $I=1/2$.
The obtained energies with various cutoffs $\Lambda$ are summarized in
Fig.~\ref{fig:energyspectrum}
and Table~\ref{table:resultsPB}.
The energy above the $\bar{D}\Lambda_{\rm c}$ threshold ($=4153.5$ MeV) is given by 
the complex value $E=E_{\rm re}-i\Gamma/2$ with 
the resonance energy $E_{\rm re}$, and the decay width $\Gamma$ for the
meson-baryon scattering states considered in this analysis. 
The real energy below 
the $\bar{D}\Lambda_{\rm c}$ threshold gives the binding energy by
subtracting the value of the $\bar{D}\Lambda_{\rm c}$ threshold.
Fig.~\ref{fig:energyspectrum} does not show the bound states with large
binding energy 
because the hadronic molecular picture is not applicable to the deeply
bound state~\cite{Ohkoda:2011vj,Yamaguchi:2013ty}.
Fig.~\ref{fig:energyspectrum} shows that the energy of states decreases
when the cutoff $\Lambda$ increases.
In large $\Lambda$ regions, the deeply bound state appears.

The cutoff parameter $\Lambda$ is fixed to reproduce the observed
pentaquarks.
We then focus on the narrow resonance  $P^+_{\rm c}(4450)$
whose significance is 12 standard deviations.
In our calculations, 
the state close to the mass of $P^+_{\rm c}(4450)$,
$4449.8\pm 1.7 \pm 2.5$ MeV.
is the $J^P=5/2^-$ state with 
the resonance energy $4428.6$ MeV in $\Lambda=1400$.
Hence, the cutoff $\Lambda$ is determined as $\Lambda=1400$,
and the $J^P$ assignment of $P^+_{\rm c}(4450)$ is $J^P=5/2^-$.
This result shows that the state corresponding to $P^+_{\rm c}(4380)$
has $J=3/2$ and the opposite parity of $P^+_{\rm c}(4450)$, 
namely $J^P=3/2^+$.
In $\Lambda=1400$ MeV, the mass of the second state of $J^P=3/2^+$ 
is $4339.7$ MeV, which
is near the mass of $P^+_{\rm c}(4380)$,  $4380\pm 8\pm 29$ MeV.
Therefore, the state obtained with $J^P=3/2^+$
corresponds to the observed pentaquark $P^+_{\rm c}(4380)$.
On the other hand, 
we find resonances other than the observed pentaquarks
in $\Lambda = 1400$ MeV.
These states are new predictions in this study.
As shown in Table~\ref{table:resultsPB},
the $J^P=3/2^-$ state has three states, whose masses are 
4136.0 MeV, 4307.9 MeV and 4348.7 MeV, respectively,
and the $J^P=3/2^+$ state has one state with the mass, 4206.7 MeV.
By contrast, the state with $J^P=5/2^+$
is absent in $\Lambda=1400$ MeV.

Our results are compared with 
those of the earlier studies on
hidden-charm molecular
states with $J^P=3/2^-$\cite{PhysRevLett.105.232001,PhysRevC.84.015202,Xiao:2013yca}.
As summarized in Table~\ref{table:comparisonwith},
the energies obtained in this study are 
slightly greater than
the results of the earlier works,
where the full-channel coupling of
$\bar{D}^{(\ast)}\Lambda_{\rm c}-\bar{D}^{(\ast)}\Sigma^{(\ast)}_{\rm c}$ 
was not considered. 
In our calculation, we find that 
the masses increase by tens of MeV
and some of states disappear
when the $\bar{D}^{(\ast)}\Lambda_{\rm c}$ channel or the states with
large orbital angular momentum are ignored, as summarized in
Table~\ref{table:channelcut}.
Specifically the $J^P=5/2^-$ state corresponding to $P^+_{\rm c}(4450)$
disappears when the analysis with the full channel coupling is not performed.

\begin{table*}[t]
 \caption{Comparison between the lowest mass of hidden-charm meson-baryon molecules
 with $I(J^P)=1/2(3/2^-)$ yielded by this work and those yielded by 
 the early
 works~\cite{PhysRevLett.105.232001,PhysRevC.84.015202,Xiao:2013yca}.
 The masses are shown in the second column in the unit of MeV.
 In this work, the value of $\Lambda$ is $1400$ MeV.
 The third column gives the channels which are considered in those works.
 }
 \label{table:comparisonwith}
 \begin{center}
  \begin{tabular}{ccc}
   \hline\hline
   Ref. 
   & Mass [MeV]& Channels\\ \hline
      This work &$4136.0$ &
	       $\bar{D}\Lambda_{\rm c}, 
 	        \bar{D}^{\ast}\Lambda_{\rm c}, 
	        \bar{D}\Sigma_{\rm c},\bar{D}\Sigma^{\ast}_{\rm c}, 
      	            \bar{D}^{\ast}\Sigma_{\rm c},
                    \bar{D}^{\ast}\Sigma^{\ast}_{\rm c}$ \\ 
   \cite{PhysRevLett.105.232001}
       & 4415& $\bar{D}^{\ast}\Sigma_{\rm c}, 
	       \bar{D}^{\ast}\Sigma^{\ast}_{\rm c}$ with only $S$-wave\\
   \cite{PhysRevC.84.015202}
       & 4454& $\bar{D}^{\ast}\Sigma_{\rm c}, 
	       \bar{D}^{\ast}\Sigma^{\ast}_{\rm c}$ with only $S$-wave\\
   \cite{Xiao:2013yca} 
       & 4334.5&
	       $J/\psi N,\bar{D}^{\ast}\Lambda_{\rm c}, 
	       \bar{D}^{\ast}\Sigma_{\rm c}, 
	       \bar{D}\Sigma^{\ast}_{\rm c},
	       \bar{D}^{\ast}\Sigma^{\ast}_{\rm c}$ 
	       \\ 
   & %&
	   & with only $S$-wave \\
   \hline\hline
  \end{tabular}
 \end{center}
\end{table*}

\begin{table}[t]
 \caption{Obtained masses with full channel coupling (Full), without
 $\bar{D}^{(\ast)}\Lambda_{\rm c}$ (w/o $\bar{D}^{(\ast)}\Lambda_{\rm
 c}$)
 and without large orbital angular
 momentum $\ell$ (w/o $\ell > 0$ or w/o $\ell > 1$)
 in $\Lambda= 1400$ MeV.
 The masses with full channel coupling are given
 in
 Table~\ref{table:resultsPB}.
 }
 \label{table:channelcut}
 \begin{center}
  \begin{tabular}{ccl}%{cccc}
   \hline\hline
   $J^P$&Channels&Mass [MeV]\\ \hline
   $3/2^-$&Full&4136.0, 4307.9, 4348.7\\
   &w/o $\bar{D}^{(\ast)}\Lambda_{\rm c}$&4278.4, 4400.4\\
   &w/o $\ell > 0$&4220.4, 4376.6\\ \hline
   $3/2^+$&Full&4206.7, 4339.7\\
   &w/o $\bar{D}^{(\ast)}\Lambda_{\rm c}$& --- \\
   &w/o $\ell > 1$&4275.3\\ \hline
   $5/2^-$&Full&4428.6\\
   &w/o $\bar{D}^{(\ast)}\Lambda_{\rm c}$& --- \\
   &w/o $\ell > 0$& ---\\
   \hline\hline
  \end{tabular}
\end{center}
\end{table}
%========================================%
\section{Summary}
\label{Sec:Summary}
%========================================%
We studied 
the hidden-charm pentaquarks as meson-baryon
molecules. 
We took into account 
the coupled channels of 
$\bar{D}^{(\ast)}\Sigma^{(\ast)}_{\rm c}$
whose thresholds are close to each other owing to the heavy quark spin symmetry.
In addition, the couplings to $\bar{D}^{(\ast)}\Lambda_{\rm c}$ near the 
$\bar{D}^{(\ast)}\Sigma^{(\ast)}_{\rm c}$ thresholds,
 and to
the states with
larger orbital angular momentum mixed by the tensor force
were considered.
Therefore, the analysis of the hidden-charm molecular systems involved
by the full coupled channel for 
$\bar{D}^{(\ast)}\Lambda_{\rm c}-\bar{D}^{(\ast)}\Sigma^{(\ast)}_{\rm
c}$, 
which had not been performed in the early works.
As for the meson-baryon interaction, the meson exchange potential was
obtained by the effective Lagrangians
that respects the heavy quark and chiral symmetries.
By solving the coupled-channel Schr\"odinger equations,
we studied 
the bound and resonant states 
in $I(J^P)=1/2(3/2^\pm)$ and $1/2(5/2^\pm)$.
The results show that the $J^P$ assignments of $P^+_{\rm c}(4380)$
and $P^+_{\rm c}(4450)$ are $3/2^+$ and $5/2^-$, respectively.
We also found new states in $J^P=3/2^\pm$.
In the molecular states obtained,
we found that 
the coupling to the $\bar{D}^{(\ast)}\Lambda_{\rm c}$ channel and 
to the state with large orbital angular momentum 
produced the attraction.
The predicted states can be sought in future experiments by
the relativistic heavy ion collision in 
LHC, 
the production via the hadron beam in J-PARC~\cite{Garzon:2015zva,Liu:2016dli,Kim:2016cxr},
the photoproduction in Jefferson Lab~\cite{Huang:2013mua,Wang:2015jsa,Huang:2016tcr} and so on.

%==========================================================
\section*{Acknowledgments}
%==========================================================
This work was supported in part by the INFN Fellowship
Programme.

%/////////////////////////////////////////////////
%                   REFERENCES
%/////////////////////////////////////////////////
\bibliographystyle{h-physrev5}
\bibliography{./reference.bib}

\begin{thebibliography}{10}

\bibitem{Aaij:2015tga}
LHCb, R.~Aaij {\em et~al.},
\newblock Phys. Rev. Lett. {\bf 115}, 072001 (2015), arXiv:1507.03414.
%%CITATION = ARXIV:1507.03414;%%

\bibitem{Aaij:2016phn}
LHCb, R.~Aaij {\em et~al.},
\newblock (2016), arXiv:1604.05708.
%%CITATION = ARXIV:1604.05708;%%

\bibitem{Aaij:2016ymb}
LHCb, R.~Aaij {\em et~al.},
\newblock (2016), arXiv:1606.06999.
%%CITATION = ARXIV:1606.06999;%%

\bibitem{Wang:2011rga}
W.~L. Wang, F.~Huang, Z.~Y. Zhang, and B.~S. Zou,
\newblock Phys. Rev. {\bf C84}, 015203 (2011), arXiv:1101.0453.
%%CITATION = ARXIV:1101.0453;%%

\bibitem{Yuan:2012wz}
S.~G. Yuan, K.~W. Wei, J.~He, H.~S. Xu, and B.~S. Zou,
\newblock Eur. Phys. J. {\bf A48}, 61 (2012), arXiv:1201.0807.
%%CITATION = ARXIV:1201.0807;%%

\bibitem{Maiani:2015vwa}
L.~Maiani, A.~D. Polosa, and V.~Riquer,
\newblock Phys. Lett. {\bf B749}, 289 (2015), arXiv:1507.04980.
%%CITATION = ARXIV:1507.04980;%%

\bibitem{Li:2015gta}
G.-N. Li, X.-G. He, and M.~He,
\newblock JHEP {\bf 12}, 128 (2015), arXiv:1507.08252.
%%CITATION = ARXIV:1507.08252;%%

\bibitem{Yang:2015bmv}
G.~Yang and J.~Ping,
\newblock (2015), arXiv:1511.09053.
%%CITATION = ARXIV:1511.09053;%%

\bibitem{Santopinto:2016pkp}
E.~Santopinto and A.~Giachino,
\newblock (2016), arXiv:1604.03769.
%%CITATION = ARXIV:1604.03769;%%

\bibitem{Hofmann:2005sw}
J.~Hofmann and M.~Lutz,
\newblock Nucl.Phys. {\bf A763}, 90 (2005), arXiv:hep-ph/0507071.
%%CITATION = HEP-PH/0507071;%%

\bibitem{PhysRevLett.105.232001}
J.-J. Wu, R.~Molina, E.~Oset, and B.~S. Zou,
\newblock Phys. Rev. Lett. {\bf 105}, 232001 (2010).

\bibitem{PhysRevC.84.015202}
J.-J. Wu, R.~Molina, E.~Oset, and B.~S. Zou,
\newblock Phys. Rev. C {\bf 84}, 015202 (2011).

\bibitem{Garcia-Recio:2013gaa}
C.~Garcia-Recio, J.~Nieves, O.~Romanets, L.~L. Salcedo, and L.~Tolos,
\newblock Phys. Rev. {\bf D87}, 074034 (2013), arXiv:1302.6938.
%%CITATION = ARXIV:1302.6938;%%

\bibitem{Xiao:2013yca}
C.~W. Xiao, J.~Nieves, and E.~Oset,
\newblock Phys. Rev. {\bf D88}, 056012 (2013), arXiv:1304.5368.
%%CITATION = ARXIV:1304.5368;%%

\bibitem{Karliner:2015ina}
M.~Karliner and J.~L. Rosner,
\newblock Phys. Rev. Lett. {\bf 115}, 122001 (2015), arXiv:1506.06386.
%%CITATION = ARXIV:1506.06386;%%

\bibitem{Uchino:2015uha}
T.~Uchino, W.-H. Liang, and E.~Oset,
\newblock Eur. Phys. J. {\bf A52}, 43 (2016), arXiv:1504.05726.
%%CITATION = ARXIV:1504.05726;%%

\bibitem{Chen:2016heh}
R.~Chen, X.~Liu, and S.-L. Zhu,
\newblock Nucl. Phys. {\bf A} (2016), arXiv:1601.03233.
%%CITATION = ARXIV:1601.03233;%%

\bibitem{Shimizu:2016rrd}
Y.~Shimizu, D.~Suenaga, and M.~Harada,
\newblock Phys. Rev. {\bf D93}, 114003 (2016), arXiv:1603.02376.
%%CITATION = ARXIV:1603.02376;%%

\bibitem{Chen:2015moa}
H.-X. Chen, W.~Chen, X.~Liu, T.~G. Steele, and S.-L. Zhu,
\newblock Phys. Rev. Lett. {\bf 115}, 172001 (2015), arXiv:1507.03717.
%%CITATION = ARXIV:1507.03717;%%

\bibitem{Wang:2015epa}
Z.-G. Wang,
\newblock Eur. Phys. J. {\bf C76}, 70 (2016), arXiv:1508.01468.
%%CITATION = ARXIV:1508.01468;%%

\bibitem{Guo:2015umn}
F.-K. Guo, U.-G. Mei{\ss}ner, W.~Wang, and Z.~Yang,
\newblock Phys. Rev. {\bf D92}, 071502 (2015), arXiv:1507.04950.
%%CITATION = ARXIV:1507.04950;%%

\bibitem{Liu:2015fea}
X.-H. Liu, Q.~Wang, and Q.~Zhao,
\newblock Phys. Lett. {\bf B757}, 231 (2016), arXiv:1507.05359.
%%CITATION = ARXIV:1507.05359;%%

\bibitem{Guo:2016bkl}
F.-K. Guo, U.-G. Mei{\ss}ner, J.~Nieves, and Z.~Yang,
\newblock (2016), arXiv:1605.05113.
%%CITATION = ARXIV:1605.05113;%%

\bibitem{Oset:1997it}
E.~Oset and A.~Ramos,
\newblock Nucl. Phys. {\bf A635}, 99 (1998), arXiv:nucl-th/9711022.
%%CITATION = NUCL-TH/9711022;%%

\bibitem{Krippa:1998us}
B.~Krippa,
\newblock Phys. Rev. {\bf C58}, 1333 (1998), arXiv:hep-ph/9803332.
%%CITATION = HEP-PH/9803332;%%

\bibitem{Hyodo:2011ur}
T.~Hyodo and D.~Jido,
\newblock Prog. Part. Nucl. Phys. {\bf 67}, 55 (2012), arXiv:1104.4474.
%%CITATION = 1104.4474;%%

\bibitem{PhysRevLett.91.262001}
Belle Collaboration, S.-K. Choi {\em et~al.},
\newblock Phys. Rev. Lett. {\bf 91}, 262001 (2003).

\bibitem{Swanson:2006st}
E.~S. Swanson,
\newblock Phys. Rept. {\bf 429}, 243 (2006), arXiv:hep-ph/0601110.
%%CITATION = HEP-PH/0601110;%%

\bibitem{Brambilla:2010cs}
N.~Brambilla {\em et~al.},
\newblock Eur. Phys. J. {\bf C71}, 1534 (2011), arXiv:1010.5827.
%%CITATION = 1010.5827;%%

\bibitem{Gamermann:2007fi}
D.~Gamermann and E.~Oset,
\newblock Eur. Phys. J. {\bf A33}, 119 (2007), arXiv:0704.2314.
%%CITATION = ARXIV:0704.2314;%%

\bibitem{Ferretti:2013faa}
J.~Ferretti, G.~Galata, and E.~Santopinto,
\newblock Phys. Rev. {\bf C88}, 015207 (2013), arXiv:1302.6857.
%%CITATION = ARXIV:1302.6857;%%

\bibitem{Ablikim:2013mio}
BESIII Collaboration, M.~Ablikim {\em et~al.},
\newblock Phys. Rev. Lett. {\bf 110}, 252001 (2013), 1303.5949.
%%CITATION = ARXIV:1303.5949;%%

\bibitem{PhysRevLett.108.122001}
Belle Collaboration, A.~Bondar {\em et~al.},
\newblock Phys. Rev. Lett. {\bf 108}, 122001 (2012).

\bibitem{Wang:2013cya}
Q.~Wang, C.~Hanhart, and Q.~Zhao,
\newblock Phys.Rev.Lett. {\bf 111}, 132003 (2013), arXiv:1303.6355.
%%CITATION = ARXIV:1303.6355;%%

\bibitem{Bondar:2011ev}
A.~E. Bondar, A.~Garmash, A.~I. Milstein, R.~Mizuk, and M.~B. Voloshin,
\newblock Phys. Rev. {\bf D84}, 054010 (2011), arXiv:1105.4473.
%%CITATION = ARXIV:1105.4473;%%

\bibitem{Ohkoda:2011vj}
S.~Ohkoda, Y.~Yamaguchi, S.~Yasui, K.~Sudoh, and A.~Hosaka,
\newblock Phys. Rev. {\bf D86}, 014004 (2012), arXiv:1111.2921.
%%CITATION = ARXIV:1111.2921;%%

\bibitem{Hosaka:2016pey}
A.~Hosaka, T.~Iijima, K.~Miyabayashi, Y.~Sakai, and S.~Yasui,
\newblock PTEP {\bf 2016}, 062C01 (2016), arXiv:1603.09229.
%%CITATION = ARXIV:1603.09229;%%

\bibitem{Isgur:1989vq}
N.~Isgur and M.~B. Wise,
\newblock Phys. Lett. {\bf B232}, 113 (1989).
%%CITATION = PHLTA,B232,113;%%

\bibitem{Isgur:1989ed}
N.~Isgur and M.~B. Wise,
\newblock Phys. Lett. {\bf B237}, 527 (1990).
%%CITATION = PHLTA,B237,527;%%

\bibitem{PhysRevLett.66.1130}
N.~Isgur and M.~B. Wise,
\newblock Phys. Rev. Lett. {\bf 66}, 1130 (1991).

\bibitem{ManoharWise200707}
A.~V. Manohar and M.~B. Wise,
\newblock {\em Heavy Quark Physics (Cambridge Monographs on Particle Physics,
  Nuclear Physics and Cosmology)}, 1 ed. (Cambridge University Press, 2007).

\bibitem{PhysRevC.40.R7}
I.~R. Afnan and B.~F. Gibson,
\newblock Phys. Rev. C {\bf 40}, R7 (1989).

\bibitem{Yasui:2009bz}
S.~Yasui and K.~Sudoh,
\newblock Phys. Rev. {\bf D80}, 034008 (2009).

\bibitem{PhysRevD.84.014032}
Y.~Yamaguchi, S.~Ohkoda, S.~Yasui, and A.~Hosaka,
\newblock Phys. Rev. D {\bf 84}, 014032 (2011).

\bibitem{PhysRevD.85.054003}
Y.~Yamaguchi, S.~Ohkoda, S.~Yasui, and A.~Hosaka,
\newblock Phys. Rev. D {\bf 85}, 054003 (2012).

\bibitem{Yamaguchi:2013ty}
Y.~Yamaguchi, S.~Ohkoda, S.~Yasui, and A.~Hosaka,
\newblock Phys. Rev. {\bf D87}, 074019 (2013), arXiv:1301.4557.
%%CITATION = ARXIV:1301.4557;%%

\bibitem{PhysRevD.72.074010}
T.~D. Cohen, P.~M. Hohler, and R.~F. Lebed,
\newblock Phys. Rev. D {\bf 72}, 074010 (2005).

\bibitem{Gamermann:2010zz}
D.~Gamermann, C.~Garc\'\i{}a-Recio, J.~Nieves, L.~L. Salcedo, and L.~Tolos,
\newblock Phys. Rev. D {\bf 81}, 094016 (2010).

\bibitem{Liang:2014kra}
W.~H. Liang, T.~Uchino, C.~W. Xiao, and E.~Oset,
\newblock Eur. Phys. J. {\bf A51}, 16 (2015), arXiv:1402.5293.
%%CITATION = ARXIV:1402.5293;%%

\bibitem{Hosaka:2016ypm}
A.~Hosaka, T.~Hyodo, K.~Sudoh, Y.~Yamaguchi, and S.~Yasui,
\newblock (2016), arXiv:1606.08685.
%%CITATION = ARXIV:1606.08685;%%

\bibitem{PhysRevD.45.R2188}
M.~B. Wise,
\newblock Phys. Rev. D {\bf 45}, R2188 (1992).

\bibitem{Falk:1992cx}
A.~F. Falk and M.~E. Luke,
\newblock Phys.Lett. {\bf B292}, 119 (1992), arXiv:hep-ph/9206241.
%%CITATION = HEP-PH/9206241;%%

\bibitem{PhysRevD.46.1148}
T.-M. Yan {\em et~al.},
\newblock Phys. Rev. D {\bf 46}, 1148 (1992).

\bibitem{Casalbuoni:1996pg}
R.~Casalbuoni {\em et~al.},
\newblock Phys. Rept. {\bf 281}, 145 (1997), arXiv:hep-ph/9605342.
%%CITATION = HEP-PH/9605342;%%

\bibitem{Bando:1987br}
M.~Bando, T.~Kugo, and K.~Yamawaki,
\newblock Phys. Rept. {\bf 164}, 217 (1988).
%%CITATION = PRPLC,164,217;%%

\bibitem{Agashe:2014kda}
Particle Data Group, K.~A. Olive {\em et~al.},
\newblock Chin. Phys. {\bf C38}, 090001 (2014).
%%CITATION = CHPHD,C38,090001;%%

\bibitem{PhysRevD.68.114001}
C.~Isola, M.~Ladisa, G.~Nardulli, and P.~Santorelli,
\newblock Phys. Rev. D {\bf 68}, 114001 (2003).

\bibitem{PhysRevD.68.054024}
W.~A. Bardeen, E.~J. Eichten, and C.~T. Hill,
\newblock Phys. Rev. D {\bf 68}, 054024 (2003).

\bibitem{Liu:2011xc}
Y.-R. Liu and M.~Oka,
\newblock Phys.Rev. {\bf D85}, 014015 (2012), arXiv:1103.4624.
%%CITATION = ARXIV:1103.4624;%%

\bibitem{Rarita:1941mf}
W.~Rarita and J.~Schwinger,
\newblock Phys. Rev. {\bf 60}, 61 (1941).
%%CITATION = PHRVA,60,61;%%

\bibitem{Hiyama:2003cu}
E.~Hiyama, Y.~Kino, and M.~Kamimura,
\newblock Prog. Part. Nucl. Phys. {\bf 51}, 223 (2003).
%%CITATION = PPNPD,51,223;%%

\bibitem{Aguilar1971_269}
J.~Aguilar and J.~M. Combes,
\newblock Communications in Mathematical Physics {\bf 22}, 269 (1971).

\bibitem{Blaslev1971_22}
E.~Balslev and J.~M. Combes,
\newblock Communications in Mathematical Physics {\bf 22}, 280 (1971).

\bibitem{Simon1972_27}
B.~Simon,
\newblock Communications in Mathematical Physics {\bf 27}, 1 (1972).

\bibitem{Prog.Theor.Phys11620061Aoyama}
S.~Aoyama, T.~Myo, K.~Kato, and K.~Ikeda,
\newblock Prog. Theor. Phys. {\bf 116}, 1 (2006).

\bibitem{Garzon:2015zva}
E.~J. Garzon and J.-J. Xie,
\newblock Phys. Rev. {\bf C92}, 035201 (2015), arXiv:1506.06834.
%%CITATION = ARXIV:1506.06834;%%

\bibitem{Liu:2016dli}
X.-H. Liu and M.~Oka,
\newblock (2016), arXiv:1602.07069.
%%CITATION = ARXIV:1602.07069;%%

\bibitem{Kim:2016cxr}
S.-H. Kim, H.-C. Kim, and A.~Hosaka,
\newblock (2016), arXiv:1605.02919.
%%CITATION = ARXIV:1605.02919;%%

\bibitem{Huang:2013mua}
Y.~Huang, J.~He, H.-F. Zhang, and X.-R. Chen,
\newblock J. Phys. {\bf G41}, 115004 (2014), arXiv:1305.4434.
%%CITATION = ARXIV:1305.4434;%%

\bibitem{Wang:2015jsa}
Q.~Wang, X.-H. Liu, and Q.~Zhao,
\newblock Phys. Rev. {\bf D92}, 034022 (2015), arXiv:1508.00339.
%%CITATION = ARXIV:1508.00339;%%

\bibitem{Huang:2016tcr}
Y.~Huang, J.-J. Xie, J.~He, X.~Chen, and H.-F. Zhang,
\newblock (2016), arXiv:1604.05969.
%%CITATION = ARXIV:1604.05969;%%

\end{thebibliography}

\end{document}